\documentclass[epj]{svjour}
\usepackage{graphics}
\begin{document}
\title{The Color Glass Condensate: an overview}
\author{Raju Venugopalan\inst{} % etc
\thanks{\emph{Present address:} Institute for Theoretical Physics II,
Univ. of Hamburg, Luruper Chaussee 49, 22761, Hamburg, Germany}%
}                     % Do not remove
\institute{Physics Department, Brookhaven National Laboratory, Upton, NY 11973, USA}
\date{Received: date / Revised version: date}
\abstract{
The Color Glass Condensate is a theory of the dynamical properties of partons in the Regge limit 
of QCD: $x_{\rm Bj}\rightarrow 0$, $Q^2 >> \Lambda_{\rm QCD}^2={\rm fixed}$ and the center of mass 
energy squared $s\rightarrow \infty$. We provide a brief introduction to the theoretical ideas underlying the Color Glass Condensate and discuss the application of these ideas to high energy scattering in QCD. }

\PACS{
      {PACS-key}{describing text of that key}   \and
      {PACS-key}{describing text of that key}
     } % end of PACS codes
%end of abstract
%
\maketitle
\section{Introduction}
\label{intro}The study of the properties of the strong interactions in the asymptotic Bjorken limit of momentum transfer squared $Q^2\rightarrow \infty$, the center of mass energy squared $s\rightarrow \infty$, and the 
Bjorken variable $x_{\rm Bj}\approx Q^2/s = {\rm fixed}$ has proved to be one of the most creative ideas in theoretical physics~\cite{Wilczek-nobel}. Relatively little work has been done in the other high energy limit, namely, 
$x_{\rm Bj}\rightarrow 0$, $s\rightarrow \infty$ and $Q^2 = {\rm fixed}$. This limit of the 
strong interactions, which we shall call the Regge limit, was studied intensively in the 60's and 
indeed led eventually to string theory. The reason these studies fell into disfavor in the strong interactions was that there 
was no small parameter in these studies (in modern parlance, $Q^2 \leq \Lambda_{\rm QCD}^2$). 

With the advent of the collider era, we can now probe a wide window of physics where $s >> Q^2 >> 
\Lambda_{\rm QCD}^2$. In fact, this "window" describes the bulk of the high energy cross-section. One therefore has finally the possibility of studying the properties of the Regge limit of the theory using 
{\it weak coupling methods}. In this limit, the hadron behaves like matter that's dense but weakly 
coupled-not dissimilar to much of condensed matter physics~\cite{IV,reviews}. 
 
 In Regge asymptotics, the number of partons increases rapidly due to QCD bremsstrahlung. This growth is described, in the leading logarithmic approximation in x, by the BFKL equation~\cite{BFKL}. 
Since the typical size of the partons in this limit is of order $1/Q^2$, the hadron becomes closely 
packed when the number of partons is of order $R^2 Q^2$. In fact, this corresponds to an occupation 
number $f\sim 1/\alpha_S$. When the density of partons is of this order, repulsive many body "recombination" and screening effects 
compete with QCD Bremsstrahlung leading to a saturation of the number of partons in the hadron's 
wavefunction~\cite{GLR,MuellerQiu,BlaizotMueller}. 
The saturation of partons of different sizes happens at different values of $x$. The scale at which this occurs is the saturation scale $Q_s(x)$--a dynamically generated 
semi-hard scale that controls the dynamics of physics in this regime of QCD.

In the language of the Operator Product Regime (OPE), the line $Q\equiv Q_s(x)$ in the $x$-$Q^2$ plane denotes the regime beyond which (when approached from high $Q^2$) higher twist effects become important. Recall that the OPE is best formulated in the Bj-limit where higher twists are 
power suppressed and can be forgotten. The opposite is true in the Regge limit. Since the number 
of twist operators grows (nearly) exponentially with the twist, the OPE quickly becomes unwieldy. Thus to describe physics in this regime we need a new organizing principle in QCD beyond the OPE.

\section{A classical effective theory for high energy QCD}
\label{sect:1}

A way out was suggested when it was realized that the physics of high parton densities could be 
formulated as a classical effective theory~\cite{MV}. When a quantum field theory is formulated on the light cone, one realizes that there is a formal Born-Oppenheimer separation between large x and small x 
modes~\cite{Susskind} which are respectively the slow and fast modes in the effective theory. Thus on the time scale of the "wee" parton small x fields, the large x partons can be viewed as static charges. 
Since these are color charges, they cannot be integrated out of the theory but must be viewed as 
sources of color charge for the dynamical wee fields. With this dynamical principle in mind, one 
can write down an effective action for wee partons in QCD at high energies. The generating functional 
of wee partons has the form
\begin{eqnarray}
{\cal Z}[j] = \int [d\rho]\, W_{\Lambda^+}[\rho]\,\left\{ {\int^{\Lambda^+} [dA] \delta(A^+) e^{iS[A,\rho]-j\cdot A} \over \int^{\Lambda^+} [dA] \delta(A^+) e^{iS[A,\rho]}}\right\} 
\label{eq:1}
\end{eqnarray}
where the wee parton action has the form
\begin{eqnarray}
& &S[A,\rho] ={-1\over 4}\,\int d^4 x \,F_{\mu\nu}^2
+ {i\over N_c}\, \int d^2 x_\perp dx^- \delta(x^-)\nonumber \\
&\times&{\rm Tr}\left(\rho(x_\perp)U_{-\infty,\infty}[A^-]\right) \, .
\label{eq:2}
\end{eqnarray}
In Eq.~\ref{eq:1}, $\rho$ is a classical color charge density (more on this shortly) of the static sources and 
$W[\rho]$ is a weight functional of sources (which sit at momenta $k^+ > \Lambda^+$: note, 
$x = k^+/P_{\rm hadron}^+$). The sources are coupled to the dynamical wee gluon fields (which in 
turn sit at $k^+ < \Lambda^+$) via the gauge invariant term~\footnote{This is not the only possible gauge 
invariant coupling. An alternative form is given in Ref.~\cite{JJV}-it can be shown to reproduce BFKL more 
efficiently.} which is the first term on the RHS of 
Eq.~\ref{eq:2}. The second term in Eq.~\ref{eq:2} is the QCD field strength tensor squared-thus the wee 
gluons are treated in full generality in this effective theory, which is formulated in the light cone gauge $A^+=0$. The source $j$ is an external source-derivatives taken with respect to this source (with the source then put to zero) generate correlation functions in the usual fashion. 

We have not justified thus far why the sources are classical. The argument for this is subtle and follows from a coarse graining of the effective action to only include modes of interest. For large nuclei, or at 
small x, the wee partons couple to a large number of sources. For a large nucleus, it can be shown 
explicitly that this source density is classical~\cite{SR}. Further, it was conjectured that the weight functional for a large nucleus was a Gaussian in the source density (corresponding to the 
quadratic Casimir operator)~\cite{MV,Kovchegov}. This was shown explicitly recently to be the correct-albeit with small corrections for SU($N_c$) coming from the $N_c-2$ higher Casimir operators~\cite{SR}.  

For a large nucleus, the variance of the Gaussian, the color charge squared per unit area $\mu_A^2$, 
proportional to $A^{1/3}$, is a large scale-and is the only scale in the effective action~\footnote{$\mu_A^2$ is simply related in the classical 
theory to the saturation scale $Q_s^2$ via the relation $Q_s^2 =\alpha_S N_c \mu_A^2 \ln(Q_s^2/\Lambda_{\rm QCD}^2)$}. Thus for $\mu_A^2 >> \Lambda_{\rm QCD}^2$, 
$\alpha_S(\mu_A^2) <<1$, and one can compute the properties of the theory in Eq.~\ref{eq:1} in weak coupling. 

By evaluating the saddle point of the action in Eq.~\ref{eq:2}, one can compute the classical distribution of gluons in the nucleus. The Yang-Mills 
equations can be solved analytically to obtain the classical field of the nucleus as a function of $\rho$: $A_{\rm cl.}(\rho)$~\cite{MV,Kovchegov,JKMW}. From the generating functional in Eq.~\ref{eq:1}, one obtains for the two point correlator, 
\begin{eqnarray}
<A A> = \int [d\rho]\,W_{\Lambda^+}[\rho]\, A_{\rm cl.}(\rho)\,A_{\rm cl.}(\rho) \, .
\label{eq:3}
\end{eqnarray}
From this expression one can determine, for Gaussian sources,  the occupation number $\phi = dN/\pi R^2/ dk_\perp^2 dy$ of wee partons in the classical field of the nucleus.
One finds that for $k_\perp >> Q_s^2$, one has the Weizs\"acker-Williams spectrum $\phi \sim Q_s^2 / k_\perp^2$, while for $k_\perp \leq Q_s$, one has a complete resummation to all 
orders in $k_\perp$, which gives $\phi\sim {1\over \alpha_S} \ln(Q_s/k_\perp)$. (The behavior at low $k_\perp$ can, more accurately, be represented 
as ${1\over \alpha_S} \Gamma(0,z)$ where $\Gamma$ is the incomplete Gamma function and $z = k_\perp^2/Q_s^2$.) A very nice expression for the classical field of the nucleus containing these two limits was presented by Triantafyllopoulos at this conference~\cite{Dionysis}. 

We are now in a position to discuss why a high energy hadron behaves like a Color Glass Condensate~\cite{IV}. The "color" is obvious since the degrees of freedom, the partons, are colored. It is a glass because the stochastic sources (frozen on time scales much larger than the wee 
parton time scales) induce a stochastic (space-time dependent) coupling between the partons under quantum evolution (to be discussed in the next section)-this is analogous to a spin glass. Finally, the matter is a condensate since the wee partons have large occupation numbers (of order 
$1/\alpha_S$) and have momenta peaked about $Q_s$. As we will discuss, these properties are enhanced by quantum evolution in x. The classical 
field retains its structure-while the saturation scale grows: $Q_s(x^\prime) > Q_s(x)$ for $x^\prime < x$. 

\section{Quantum evolution {\it a la} JIMWLK and BK}
\label{sect:2}

Small fluctuations about the effective action in Eq.~\ref{eq:2} were first considered in Ref.~\cite{AJMV}. It was discovered that these 
gave large corrections of order $\alpha_S\ln(1/x)$. In particular, this suggested that the Gaussian weight functional was fragile under quantum evolution of the sources~\footnote{We will return to this point in our discussion of the Cronin effect 
in Deuteron-Gold collisions at RHIC.}. A Wilsonian renormalization group (RG) approach was developed to systematically 
treat these corrections~\cite{JIMWLK}. The basic ingredients of this approach are as follows. Begin with the generating functional in 
Eq.~\ref{eq:1} at some $\Lambda^+$, with an initial source distribution $W[\rho]$. Perform small fluctuations about the classical saddle 
point of the effective action, integrating out momentum modes in the region ${\Lambda^\prime}^+ < k^+ < \Lambda^+$,  ensuring that 
${\Lambda^\prime}^+$ is such that $\alpha_S \ln(\Lambda^+/{\Lambda^\prime}^+) << 1$. The action reproduces itself at the new 
scale ${\Lambda^\prime}^+$, albeit with a charge density $\rho^\prime = \rho + \delta \rho$, and $W_{\Lambda^+}
[\rho]\longrightarrow W_{{\Lambda^\prime}^+}[\rho^\prime]$. The change of the weight functional $W[\rho]$ with x is described by 
the JIMWLK- non-linear RG equation~\cite{JIMWLK} which we shall not  write explicitly here. 

The JIMWLK equations form an infinite hierarchy (analogous to the BBGKY hierarchy in statistical mechanics) of 
ordinary differential equations for the gluon correlators $<A_1 A_2 \cdots A_n>_Y$, where $Y= \ln(1/x)$ is the rapidity. The expectation 
value of an operator ${\cal O}$ is defined to be 
\begin{eqnarray}
<{\cal O}>_Y = \int [d\alpha] {\cal O}[\alpha] W_Y[\alpha] \,,
\label{eq:4}
\end{eqnarray}
where $\alpha = {1\over \nabla_\perp^2}\,\rho$. The corresponding JIMWLK equation for this operator is 
\begin{eqnarray}
{\partial <{\cal O}[\alpha]>_Y\over \partial Y} &=& < {1\over 2}\,\int_{x_\perp,y_\perp}\,{\delta \over \delta \alpha_Y^a(x_\perp)}\,\chi_{x_\perp,y_\perp}^{ab}[\alpha]\nonumber \\
& &{\delta \over \delta \alpha_Y^b (y_\perp)} {\cal O}[\alpha]>_Y \,.
\label{eq:5}
\end{eqnarray}
$\chi$ here is a non-local object expressed in terms of path ordered (in rapidity) Wilson lines of $\alpha$~\cite{IV}. This equation is analogous to a
(generalized) functional  Fokker-Planck equation, where $Y$ is the "time" and $\chi$ is a generalized diffusion coefficient. This equation illustrates 
the stochastic properties of operators in the space of gauge fields at high energies. For the gluon density, which is proportional to a two-point function 
$<\alpha^a (x_\perp) \alpha^b (y_\perp)>$, one recovers the BFKL equation in the limit of low parton densities.

As mentioned, the JIMWLK equations are master equations for n-point correlators.  Two point correlators of Wilson lines are proportional to 
4-point correlators and so on. The theory is conformal so it is not inconceivable that it is exactly solvable but this has not been done thus far. Preliminary numerical solutions have been obtained recently~\cite{RW} but much work remains in that 
direction. There is a mean field solution deep in the saturation regime~\cite{IancuMcLerran} where one can show that the weight functional is a Gaussian-albeit a non local Gaussian with a variance proportional to $k_\perp^2$ for $k_\perp^2 < Q_s^2$. 

In the limit of large $N_c$ and large A ($\alpha_S^2 A^{1/3} >> 1$), one can show that the hierarchy closes for the two point correlator of 
Wilson lines since  the expectation value of the product of traces of Wilson lines factorizes into the product of the expectation values of the 
traces:
\begin{equation}
<{\rm Tr}(V_x V_z^\dagger) {\rm Tr}(V_z V_y^\dagger)> \longrightarrow <{\rm Tr}(V_x V_z^\dagger)>\,<{\rm Tr}(V_z V_y^\dagger)> \, ,
\label{eq:6}
\end{equation}
where $_x={\cal P}\exp\left(\int dz^- \alpha^a(z^-,x_\perp) T^a\right)$. Here ${\cal P}$ denotes path ordering in $x^-$ and $T^a$ is the SU(3) 
generator in the adjoint representation. In Mueller's dipole picture~\footnote{See also Ref.~\cite{NZ}.}, the cross-section for a dipole 
scattering off a target can be expressed in terms of these 2-point dipole operators as~\cite{Mueller2}
\begin{equation}
\sigma_{q\bar q N} (x, r_\perp) = 2\, \int d^2 b\, \, {\cal N}_Y (x,r_\perp,b) \, ,
\label{eq:7}
\end{equation}
where ${\cal N}_Y$, the imaginary part of the forward scattering amplitude, is defined to be ${\cal N}_Y= 1 - {1\over N_c}<{\rm Tr}(V_x V_y^\dagger)>_Y$. 
Note that the size of the dipole, ${\vec r}_\perp = {\vec x}_\perp - {\vec y}_\perp$ and ${\vec b} = ({\vec x}_\perp + {\vec y}_\perp)/2$. The JIMWLK equation for the two point Wilson correlator is identical in the 
large A, large $N_c$ mean field limit to an equation derived independently by Balitsky and Kovchegov-the Balitsky-Kovchegov equation~\cite{BK}, 
which has the operator form
\begin{equation}
{{\partial {\cal N}_Y}\over \partial Y} = {\bar \alpha_S}\, {\cal K}_{\rm BFKL} \otimes \left\{ {\cal N}_Y - {\cal N}_Y^2\right\} \, .
\label{eq:8}
\end{equation}
Here ${\cal K}_{\rm BFKL}$ is the well known BFKL kernel. When ${\cal N} << 1$, the quadratic term is negligble and one has BFKL growth of the number of dipoles; when ${\cal N}$ is close to unity, 
the growth saturates. The approach to unity can be computed analytically~\cite{LevinTuchin}. The B-K equation is the simplest equation including
both the Bremsstrahlung responsible for the rapid growth of amplitudes at small x as well as the repulsive many 
body effects that lead to a saturation of this growth. 

A saturation condition which fixes the amplitude at which this change in behavior is significant, say ${\cal N} =1/2$, determines the saturation 
scale. One obtains  $Q_s^2 = Q_0^2 \exp( \lambda Y)$, where $\lambda = c \alpha_S$ with $c\approx 4.8$. The saturation condition affects the overall normalization of this scale but does not affect the power $\lambda$. In fixed coupling, the power $\lambda$ is large and there are large pre-asymptotic corrections to this relation-which die off only slowly as a function of $Y$. BFKL running coupling effects change the behavior of the saturation scale completely-one goes smoothly at large $Y$ to $Q_s^2 = Q_0^2 \exp(\sqrt{2b_0 c(Y + Y_0}))$ where $b_0$ is the coefficient of the one-loop QCD 
$\beta$-function. The state of the art computation of $Q_s$ is the work of Triantafyllopoulos, who obtained $Q_s$ by solving 
NLO-resummed BFKL in the presence of an absorptive boundary (which corresponds to the CGC)~\cite{Dionysis2}. The pre-asymptotic effects 
are much smaller in this case and the coefficient $\lambda\approx 0.25$ is very close to the value extracted from saturation model fits to the 
HERA data~\cite{GBW1}.

No analytical solution of the BK equation in the entire kinematic region but there have been several numerical studies at both fixed and running 
coupling~\cite{GSM,Armesto,Albacete}. These studies suggest that the solutions have a soliton like structure and that the saturation scale has the 
behavior discussed here. Geometrical scaling of solutions is seen for a wide window in rapidities. Running coupling effects, as suggested, are 
important and make the results of the computations more physically plausible.

The soliton like structure is no accident, as was discovered by Munier and Peschanski~\cite{MunierPeschanski} who noticed that the 
BK-equation, in a diffusion approximation, bore a formal analogy to the FKPP equation describing the propagation of unstable non-linear wavefronts~\cite{FKPP}. In addition, the full BK-equation lies in the universality class of the FKPP equation enabling one to extract the 
universal properties of these equations (for instance the leading pre-asymptotic terms in the expression for the saturation scale). 
The power of this analogy was made manifest~\cite{IMMunier} when it was realized that a stochastic generalization of the FKPP equation-the sFKPP equation-could provide insights into impact parameter dependent fluctuations~\cite{MuellerShoshi} in high energy QCD beyond the BK-equation. This 
is a very active area of research now, with several groups hunting for the Pomeron loops responsible for these fluctuations. The (rapidly evolving) 
state of the art of this subject is discussed in the talk by Edmond Iancu in these proceedings~\cite{Edmond}. 

To summarize, the Color Glass Condensate is a weak coupling effective theory describing the properties of hadron wavefunctions in QCD at high energies. Renormalization group equations-the JIMWLK 
equations-describe the behavior of multi-parton correlations in the hadron wavefunction as a function of rapidity. The theory has stochastic features 
closely analogous to the propagation of unstable non-linear wave fronts in statistical mechanics. Recent work~\cite{Edmond} is focused on trying to understand possible corrections beyond JIMWLK at low parton densities-which may be responsible for Pomeron loops. We now turn to the applications of this theory to hadronic scattering.

\section{Hadronic scattering and $k_\perp$ factorization in the Color Glass Condensate}
\label{sect:3}

Collinear factorization is the perturbative QCD mechanism to compute hard scattering. For instance, the cross-section in pp-collisions for di-jets with 
invariant mass $M^2\sim s >> \Lambda_{\rm QCD}^2$, is a convolution of  structure functions from each of the nucleons (evaluated at the scale $M^2$) times the probability that collinear partons with $k_\perp=0$ from the nuclei scatter to produce the di-jet. The structure functions are universal 
since they can be extracted from one set of experiments and input into another. Factorization theorems prove that this universality holds modulo 
power corrections in $M^2$. At collider energies, a new window opens up where $\Lambda_{\rm QCD}^2 << M^2 << s$. In principle, cross-sections 
in this window can be computed in the collinear factorization language-however, one needs to sum up large logarithmic corrections in $s/M^2$. 
An alternative formalism is that of $k_\perp$-factorization~\cite{CCH,CE}, where one has a convolution of $k_\perp$ dependent ``un-integrated" 
gluon distributions from the two hadrons with the hard scattering matrix. In this case, the in-coming partons from the wavefunctions have non-zero 
$k_\perp$. It was suggested by Levin et al.~\cite{LRSS} that at high energies the typical $k_\perp$ is the saturation scale $Q_s$. The rapidity dependence of the unintegrated distributions is given by the BFKL equation. However, unlike the structure functions, it has not been proven that 
these unintegrated distributions are universal functions. 

At small x, both the collinear factorization and $k_\perp$ factorization limits can be understood in a systematic way in the framework of the 
Color Glass Condensate. Rather than a convolution of probabilities, one has instead a collision of classical gauge fields. The expectation value of 
an operator ${\cal O}$ can be computed as 
\begin{eqnarray}
<{\cal O}>_Y = \int [d\rho_1]\, [d\rho_2]\, W_{x_1}[\rho_1]\, W_{x_2}[\rho_2]\, {\cal O}(\rho_1,\rho_2) \, ,
\label{eq:9}
\end{eqnarray}
where $Y = \ln(1/x_F)$ and $x_F = x_1 - x_2$. All operators at small $x$ can be computed in the background 
classical field of the nucleus at small $x$. Quantum 
information, to leading logarithms in $x$, is contained in the source functionals $W_{x_1 (x_2)}[\rho_1(\rho_2)]$. The operator 
${\cal O}$ can be expressed in terms of gauge fields $A^\mu[\rho_1,\rho_2](x)$. 

Inclusive gluon production in the CGC is computed by solving the Yang-Mills equations 
$[D_\mu,F^{\mu\nu}]^a = J^{\nu,a}$, where 
\begin{eqnarray}
J^\nu = \rho_{1}\,\delta(x^-)\delta^{\nu +} + \rho_{2}\,\delta(x^+)\delta^{\nu -} \, .
\label{eq:10}
\end{eqnarray}
with initial conditions given by the Yang-Mills fields of the two nuclei before the collision. These are obtained 
self-consistently by matching the solutions of the Yang-Mills equations on the light cone~\cite{KMW}. The initial conditions 
are determined by requiring that singular terms in the matching vanish. Since we have argued in Section~\ref{sect:2} that we can {\it compute} the 
Yang-Mills fields in the nuclei before the collision, the classical problem is in principle completely solvable. Quantum corrections not 
enhanced by powers of $\alpha_S\ln(1/x)$ can be included systematically. The terms so enhanced are absorbed into the weight functionals 
$W[\rho_{1,2}]$.  

Hadronic scattering in the CGC can therefore be studied through a systematic power counting in the density of sources in powers of $\rho_{1,2}/k_{\perp;1,2}^2$. This power counting in fact is more relevant at high 
energies than whether the incoming projectile is a hadron or a nucleus. In addition, 
one can begin to study the applicability of both collinear and $k_\perp$ factorization at small $x$ in this 
approach. 

\subsection{Gluon and quark production in the dilute/pp regime: ($\rho_{p1}/k_\perp^2\,\rho_{p2}/k_\perp^2 <<1$)}

The power counting here is applicable either to a proton at small x, or to a nucleus (whose parton density 
at high energies is enhanced by $A^{1/3}$) at large transverse momenta. The 
relevant quantity here is $Q_s$, which, as one may recall, is enhanced both for large $A$ and small $x$. 
So as long as $k_\perp >> Q_s >> \Lambda_{\rm QCD}$, one can consider the proton or nucleus as being dilute. 

To lowest order in $\rho_{p1}/k_\perp^2$ and $\rho_{p2}/k_\perp^2$, one can compute inclusive gluon production analytically. 
This was first done in the $A^\tau=0$ gauge ~\cite{KMW} and subsequently in the Lorentz gauge $\partial_\mu A^\mu=0$~\cite{KovRischke}.  At large transverse momenta, $Q_s << k_\perp$, the scattering can be expressed in a $k_\perp$-factorized 
 form. The inclusive cross-section is expressed as the product of 
 two unintegrated ($k_\perp$ dependent) distributions times the matrix element for the scattering. 
The comparison of this result to the collinear pQCD $gg\rightarrow gg$ process and the $k_\perp$ factorized $gg\rightarrow g$ was performed in Ref.~\cite{GyulassyMcLerran}.  At this order, the result is equivalent 
to the perturbative QCD result first derived by Gunion and Bertsch~\cite{GunionBertsch}. This 
result for gluon production is substantially modified, as we shall discuss shortly, by high parton density effects in the nuclei.  

$k_\perp$ factorization is a good assumption at large momenta for quark pair-production. This was worked out in the CGC approach by Fran\c cois Gelis and myself~\cite{FR}. The result for inclusive quark pair production can  be expressed in $k_\perp$ factorized form as
\begin{eqnarray}
& &\frac{d\sigma_1}{dy_p dy_q d^2p_\perp d^2q_\perp}
\propto
\int\frac{d^2 k_{1\perp}}{(2\pi)^2}\frac{d^2 k_{2\perp}}{(2\pi)^2}\delta(k_{1\perp}+k_{2\perp}-p_\perp-q_\perp)\nonumber \\
& &\times \varphi_1(k_{1\perp}) \varphi_2(k_{2\perp})
\frac{{\rm Tr}\,\left(\left|m^{-+}_{ab}(k_1,k_2;q,p)\right|^2\right)}
{k_{1\perp}^2 k_{2\perp}^2} \; ,
\label{eq:11}
\end{eqnarray}
where $\phi_1$ and $\phi_2$ are the unintegrated gluon distributions in the projectile and target respectively (with the gluon distribution defined as $xG(x,Q^2) = \int_0^{Q^2} d(k_\perp^2)\, \phi(x,k_\perp)$). The matrix element ${\rm Tr}\,\left(\left |m^{-+}_{ab}(k_1,k_2;q,p)\right|^2\right)$ is identical to the result derived in the $k_\perp$--factorization approach~\cite{CCH,CE}.  In the limit 
$|\vec{k_{1\perp}}|\,,|\vec{k_{2\perp}}|\rightarrow 0$, $\frac{{\rm Tr}\,\left(\left|m^{-+}_{ab}(k_1,k_2;q,p)\right|^2\right)}
{k_{1\perp}^2 k_{2\perp}^2}$ is well defined--after integration over the azimuthal angles in Eq.~\ref{eq:11}, one obtains the usual matrix 
element $|{\cal M}|_{gg\rightarrow q\bar q}^2$, recovering the lowest order pQCD collinear factorization result.  

\subsection{\bf Gluon and quark production in the semi-dense/pA region ($\rho_{p}/k_\perp^2<<1\,\rho_{A}/k_\perp^2\sim 1$).}

The power counting here is best applicable to asymmetric systems such as proton-nucleus collisions, which 
naturally satisfies the power counting for a wide range of energies. Of course, as one goes to extremely high 
energies, it is conceivable that the parton density locally in the proton can become comparable to that in the 
nucleus. In the semi-dense/pA case, one solves the Yang--Mills equations $[D_\mu,F^{\mu\nu}]=J^\nu$ with the light cone sources 
$J^{\nu,a}=\delta^{\nu+}\,\delta(x^-)\,\rho_p^a(x_\perp)+\delta^{\nu -}\,\delta(x^+)\,\rho_A^a(x_\perp)$,
to determine the gluon field produced-to lowest order in the proton source density and to all orders in the  
nuclear source density.  The inclusive gluon production cross-section, in this framework, was first computed by Kovchegov and Mueller~\cite{KovMueller} and shown to be $k_\perp$ factorizable in Ref.~\cite{KKT}. In Ref.~\cite{BFR1}, the gluon field produced in pA collisions was computed explicitly in Lorentz gauge $\partial_\mu A^\mu=0$. . 
Our result is exactly equivalent to that of Dumitru \& McLerran in $A^\tau=0$ gauge~\cite{DumitruMclerran}. The well known 
``Cronin" effect is obtained in our formalism and can be simply understood in terms of the multiple scattering of a parton from the projectile 
with those in the target. The Cronin effect and its evolution with rapidity will be discussed in the next section.

Quark production in p/D-A collisions can be computed with the gauge field in Lorentz gauge~\cite{BFR2}. The field is decomposed into the sum of `regular' terms and 'singular' terms; the latter contain $\delta(x^+)$. The regular terms are the cases where a) a gluon from the proton interacts with the nucleus and produces a $q\bar q$-pair outside, b) the gluon produces the pair which then scatters off the 
nucleus. Naively, these would appear to be the only possibilities 
in the high energy limit where the nucleus is a Lorentz contracted 
pancake. However, in the Lorentz gauge, one has terms identified with the singular terms in the gauge field which correspond to the case where the quark pair is both produced and re-scatters in the 
nucleus! 

Our result for quark pair production were discussed by Gelis and Fujii~\cite{GelisFujii} at this conference. Related work for single quark production 
was also discussed by Tuchin~\cite{Tuchin}~\footnote{In addition, see related work in Ref.~\cite{KopRauf}-
for a recent review of $k_\perp$ factorization in heavy quark production, see 
Ref.~\cite{GoncalvesMachado}.}. Unlike gluon production, neither quark pair-production nor single quark production is strictly $k_\perp$ factorizable. 
The pair production cross-section can however still be written in $k_\perp$ factorized form as a product of the unintegrated gluon distribution 
in the proton times a sum of terms with three unintegrated distributions,  $\phi_{g,g}$, $\phi_{q\bar q, g}$ and $\phi_{q\bar q,q\bar q}$. These are respectively proportional to 2-point, 3-point and 4-point correlators of the Wilson lines we discussed previously. For instance, the distribution $\phi_{q\bar q, g}$ is the product of fundamental Wilson lines coupled to a $q\bar q$ pair in the amplitude and adjoint Wilson lines coupled to a gluon in the complex conjugate amplitude. For large transverse momenta 
or large mass pairs, the 3-point and 4-point distributions collapse to the unintegrated gluon distribution, and we recover the previously discussed $k_\perp$-factorized result for pair production in the dilute/pp-limit. Single quark distributions are straightforwardly obtained and depend only on 
the 2-point quark and gluon correlators and the 3-point correlators. 
For Gaussian sources, as in the MV-model, these 2-,3- and 4-point 
functions can be computed exactly as discussed in Ref.~\cite{BFR2} and in Ref.~\cite{GelisFujii}. 

{\it The results for gluon and quark production in p/D-A collisions, coupled with the previous results for 
inclusive and diffractive~\cite{GelisJamal,Kopeliovich,KT,KovWied} distributions in DIS suggest an important new paradigm. At small $x$ in DIS and hadron colliders, previously interesting observables such as quark and gluon structure functions are 
no longer the right observables to capture the relevant physics. Instead they should be replaced by these dipole and 
multipole correlators of Wilson lines that seem ubiquitous in all high energy processes and are similarly gauge invariant 
and process independent. The renormalization group running of these operators are a powerful and sensitive harbinger of new physics.}

\subsection{\bf Gluon and quark production in the dense/AA region ($\rho_{A1}/k_\perp^2 =\rho_{A2}/k_\perp^2\sim 1$).}

In nucleus-nucleus collisions, $\rho_{1,2}/k_\perp^2\sim 1$. There is no small expansion parameter and one has thus far not been able to compute particle production analytically in the CGC.  
Unlike gluon production in the pp and pA cases, $k_\perp$-factorization breaks down in the AA-case~\cite{AR,Balitsky2}. $k_\perp$ factorization breaking terms are O(1) and there 
are a large number of these. This is because the classical field comes in with a factor $1/g$-thus each insertion on the gluon is of order O(1). A significant consequence is that one cannot factor the quantum evolution of the initial 
wavefunctions into unintegrated gluon distributions unlike the pA case.

Nevertheless, there is a systematic way to 
include small x effects in the AA case. The problem of nuclear collisions is well defined in weak coupling and can be solved numerically~\cite{AR,AYR,Lappi}.
The numerical simulations thus far assume Gaussian initial conditions as in the MV model. These are good initial conditions for central Gold-Gold collisions at RHIC where the typical $x$ is of order $10^{-2}$. They are not good initial conditions at the LHC where the 
typical $x$ at central rapidities will be at least an order of magnitude lower. In that case, one has to use solutions of 
JIMWLK RG equations~\cite{RW}.
The numerical lattice formalism of Ref.~\cite{AR} is ideal for computing particle production in the forward light cone by matching the Wilson lines from 
each of the nuclei on the light cone. 

We  restrict ourselves to discussing numerical solutions with Gaussian initial conditions. The saturation 
scale $Q_s$ (which is an input in the numerical solutions in this approximation) and the nuclear radius $R$ are the 
only parameters in the problem. 
The energy and number respectively of gluons released in a heavy ion collision of identical nuclei can therefore be simply expressed as
\begin{eqnarray}
{1\over \pi R^2}\,{dE\over d\eta} &=& {c_E\over g^2}\,Q_s^3 \,,\nonumber \\
{1\over \pi R^2}\,{dN\over d\eta} &=& {c_N\over g^2}\, Q_s^2 \, ,
\label{eq:12}
\end{eqnarray}
where (up to $10\%$ statistical uncertainity) we compute numerically $c_E=0.25$ and $c_N=0.3$. Here $\eta$ is the space-time 
rapidity. 

The number distributions of gluons can also be computed in this approach. Remarkably, one finds that a) the 
number distribution is infrared finite, and b) the distribution 
is well fit by a massive Bose-Einstein distribution for $k_\perp/Q_s < 1.5$ GeV with a ``temperature" of $\sim 0.47 Q_s$ 
and by the perturbative distribution $Q_s^4/k_\perp^4$ for $k_\perp/Q_s > 1.5$. 

\section{What the CGC tells us about matter produced in D-A and A-A collisions at RHIC}
\label{sect:4}

Gluon distributions computed in pA collisions in the MV model exhibit the Cronin effect~\cite{DumitruJamal,GelisJamal}. One can show that this 
is exactly equivalent to the Glauber picture where partons from the proton acquire transverse momenta from multiple scattering off partons in the 
target~\cite{Accardi,BFR1}. However, unlike the Glauber picture, quantum evolution in the CGC (see the discussion in Section~\ref{sect:2}) predicts that the Gaussian approximation breaks down completely when the $x_2$ in the target is such that $\ln(1/x_2)\sim 1/\alpha_S$. In other words, as one 
produces gluons further and further forward in the proton fragmentation region (recall $x_F = x_1 - x_2$), the Glauber picture should break down. 
Indeed, that is precisely the trend that is observed in the RHIC Deuteron-Gold experiments~\cite{BRAHMS1}. The rapid depletion of the Cronin effect 
is likely due to the onset of BFKL evolution, while the subsequent saturation of this trend reflects the onset of saturation effects~
\cite{KKT,Albacete,BKW,JYR,BFR1,IIT}. Further, there is  a natural explanation for the dramatic inversion of the centrality dependence that one observes in the RHIC data-
it arises due to the onset of BFKL anomalous dimensions-crudely put, the nuclear Bremsstrahlung spectrum changes from $Q_s^2/k_\perp^3 \longrightarrow Q_s/k_\perp$. Finally, an additional piece of evidence that can be adduced in support of the CGC picture is the broadening of 
azimuthal correlations~\cite{KLM2} for which preliminary data now exists from the STAR collaboration~\cite{STAR}. These ideas can be tested 
conclusively in photon and di-lepton production in D-A collisions at RHIC~\cite{BMS,Jamal,Betemps}.

We have learnt several things from applying the CGC to D-A collisions. Firstly, the Gaussian MV model works at mid-rapidities $x\sim 10^{-2}$-quantum 
evolution {\it a la} BFKL is not significant at these values of x. The MV model is therefore a good model of the initial conditions for A-A collisions at RHIC. (More on this to come shortly.) The MV model is not a good model as 
one goes forward in the Deuteron direction-at small x's of $x\sim 10^{-3}$ or lower. Quantum evolution effects, seen explicitly in solutions of the 
B-K equation-are important. They will therefore provide the initial conditions for heavy ion collisions at the LHC.

The MV model when applied to heavy ion collisions correctly predicted the initial multiplicity at RHIC~\cite{AR}. It was also remarkably successful in 
explaining rapidity distributions and the centrality dependence of multiplicities~\cite{KLN}. However, it soon became clear that the CGC alone was 
not sufficient to explain the RHIC data since a) it could not explain the RHIC $v_2$ data~\cite{AYR} and b) it predicted a suppression in D-A 
collisions at RHIC (the MV model notwithstanding) which disagreed with the RHIC data~\cite{KLM}. This failure of the CGC (here meaning 
quantum evolution as opposed to the MV model which has no evolution) thus strongly suggested that final state interactions are important at 
RHIC-which corroborates the remarkable success of hydrodynamic models.

Why do predictions of bulk features-the multiplicity~\cite{AR} and rapidity and centrality dependence~\cite{KLN} do so well then? If hydrodynamic 
behavior sets in early, and viscous effects are small, the bulk features from the initial conditions will be preserved by hydrodynamic flow. This is seen 
in the hydrodynamic simulations of Hirano and Nara~\cite{HiranoNara}. Thus one has the beginnings of a consistent phenomenological picture-though 
many puzzles remain. We don't understand why thermalization is early (more on that in the next section) or have a quantitative understanding of 
what the viscous corrections are. In general, we don't have a good understanding of the properties of the strongly interacting quark gluon plasma-
better data and better calculations will be helpful.

The RHIC data on the multiplicity (approximately 1000 hadrons in one unit of rapidity) and transverse energy 
(approximately $500$ GeV for central rapidities) of produced hadrons combined with Eq.~\ref{eq:12} place strong constraints on what $Q_s$ can be.  If $Q_s$ is too small, we find, absurdly, that the initial transverse energy is 
less than the final measured transverse energy. If $Q_s$ is too large, we find that 
the initial multiplicity of gluons is greater than the final multiplicity of hadrons.  While there is no obvious 
theorem that prohibits the initial gluon multiplicity being greater than the final hadron multiplicity, such a situation is unlikely in all 
statistical/hydrodynamic scenarios of the 
RHIC collisions. These constraints therefore allow us to place the bound that~\cite{AYR2}. 
\begin{equation}
1.3 < Q_s < 2\,\, {\rm GeV}
\label{eq:42}
\end{equation}
This bound is consistent with an $A^{1/3}$ extrapolation of the Golec-Biernat--Wusthoff fit of $Q_s^2$ 
to the HERA  data~\cite{GBW1}. A simple extrapolation gives $Q_s\approx 1.4$ GeV.  

\section{Thermalization: from CGC to QGP}
\label{sect:5}

The transition to the QGP from the CGC remains as an outstanding theoretical problem.  Due to the rapid expansion of the system, the occupation number of modes falls well below one on time scales of order $1/Q_s$. From these times onwards, one expects the 
canonical classical approach to break down-well before thermalization.  On the other hand, for elliptic flow from hydrodynamics to be significant, the conventional wisdom is that thermalization should set in early. A necessary condition is that momentum distributions should be isotropic. The CGC initial conditions are very anisotropic with $<p_\perp> \sim Q_s$ and $<p_z>\sim 0$. How does this isotropization take place? All estimates of final state re-scattering of partons formed 
from the melting CGC, both from $2\rightarrow 2$ 
processes~\cite{Mueller6} and $2\rightarrow 3$ processes~\cite{BMSS} suggest thermalization takes 
longer than what the RHIC collisions seem to suggest-in the latter case, $\tau_{\rm thermal}\sim {1\over \alpha_S^{13/5}}\,
{1\over Q_s}$, which at RHIC energies gives $\tau_{\rm thermal}\sim 2-3$ fm. 

Recently, it has been suggested that collective instabilities~\cite{Stan}, analogous  to the well known Weibel instabilities 
in plasma physics, can speed up themalization~\cite{AML,RS1}. Starting from very anisotropic (CGC-like) initial conditions, these instabilities drive the system to isotropy on very short time scales, of order $1/Q_s$ in some estimates. 
What is the relation of this language of instabilities and that of our classical field simulations? One possibility is that our particular initial conditions, the non-linearities of the fields and the rapid expansion of the system kill the growth of instabilities. Another intriguing possibility is that small violations of boost invariance provide the the seeds 
for the instabilities. Further, to properly study thermalization, one should better understand the interaction of high momentum 
(particle) and low momentum (field) degrees of freedom and their evolution. This leads to a real time renormalization group 
description~\cite{Boyanovsky}. 

An equally interesting problem is that of chemical equilibration. At high energies, the initial state in a heavy ion collision is dominated by gluons. 
Are quarks produced in sufficient numbers for the system to reach chemical equilibrium (where the ratio of gluons to quarks is expected to 
be $32/21 N_f$)? One would expect, in weak coupling, that the production of quarks to 
be suppressed. However, since the fields from the CGC are of order $1/g$, strong fields could 
drive the system to chemical equilibrium. First steps have been taken to study this problem~\cite{Dietrich,FTK} which involves numerically 
solving the Dirac equation in the background field of the two nuclei. One 
expects further progress on this problem in the near future.

\section{Open Issues in the CGC}
\label{sect:6}
The CGC is a framework to think about problems in high energy QCD.  There are many loose ends. A topic of much excitement among theorists recently is whether there 
are contributions beyond the JIMWLK equations-in particular those that generate "Pomeron loops". These contributions are likely at low to moderate parton densities where impact parameter fluctuations are large. This topic is addressed in the talk by Iancu~\cite{Edmond}. We addressed the issue of $k_\perp$-factorization and why "dipole" and "multipole" operators may be more relevant variables at high energies than structure functions. Can one derive factorization theorems in this framework analogous to those derived previously for Collinear Factorization? Turning to phenomenology, we have the beginnings of a consistent phenomenological picture of the CGC and the QGP in D-A and A-A collisions. For this to become a quantitative science, we need to understand the problem of thermalization from first principles in QCD. It is a difficult task but by no means an impossible one.

\section*{Acknowledgements}
This work was supported in part  by DOE Contract No. DE-AC02-98CH10886 and by a research 
grant from the Alexander Von Humboldt Foundation. I would like to further acknowledge the kind hospitality of the Institute for 
theoretical physics at Bielefeld University and the Institute for theoretical physics-II, at Hamburg University.

\end{document}